\documentclass[12pt,preprint]{aastex}

\newcommand\spi{{\em SPITZER}}%
\newcommand\igrj{IGR J16318--4848}%
\newcommand\gx{GX 301--2}%
\newcommand\integ{{\em INTEGRAL}}%
\newcommand\msun{$M_{\odot}$}%
\newcommand\av{$A_{\rm V}$}%
\newcommand\nh{$N_{\rm H}$}%
\newcommand\hh{$\rm H_{\rm 2}$}%
\newcommand\necubic{$N_{\rm e}$}%
\newcommand\cmcubic{$\rm cm^{\rm -3}$}%
\newcommand{\niii}{[\ion{Ni}{2}]}
\newcommand{\neii}{[\ion{Ne}{2}]}
\newcommand{\neiii}{[\ion{Ne}{3}]}
\newcommand{\feii}{[\ion{Fe}{2}]}
\newcommand{\feiii}{[\ion{Fe}{3}]}
\newcommand{\siii}{[\ion{S}{3}]}
\newcommand{\silii}{[\ion{Si}{2}]}

\slugcomment{Submitted: \today}

\shorttitle{\spi\ Mid-Infrared Spectroscopy of Highly-Obscured X-ray Binaries}
\shortauthors{Moon et al.}

\begin{document}

\title{The Rich Mid-Infrared Environments of Two Highly-Obscured
  X-ray Binaries: \spi\ Observations of \igrj\ and \gx }

\author{Dae-Sik Moon\altaffilmark{1},
David L. Kaplan\altaffilmark{2},
William T. Reach\altaffilmark{3},
Fiona A. Harrison\altaffilmark{4},
Jeong-Eun Lee\altaffilmark{5},
Peter G. Martin\altaffilmark{6} }
\altaffiltext{1}{Department of Astronomy and Astrophysics, University of Toronto, Toronto, ON M5S 3H4, Canada; moon@astro.utoronto.ca}
\altaffiltext{2}{Pappalardo Fellow, Kavli Institute for Astrophysics
  \& Space Research and Department of Physics, Massachusetts Institute of Technology, Cambridge, MA 02139; dlk@space.mit.edu}
\altaffiltext{3}{Infrared Processing Analysis Center, California Institute of Technology, MS 220-6, Pasadena, CA 91125; reach@ipac.caltech.edu}
\altaffiltext{4}{Space Radiation Laboratory, California Institute of Technology, MC 220-47, Pasadena, CA 91125; fiona@srl.caltech.edu}
%\altaffiltext{5}{Hubble Fellow, Department of Physics and Astronomy, University of California, Los Angeles, CA 90095}
\altaffiltext{5}{Department of Astronomy and Space Science, Sejong University, Seoul 143-747, Korea; jelee@sejong.ac.kr}
\altaffiltext{6}{Canadian Institute for Theoretical Astrophysics, University of Toronto, Toronto, ON M5S 3H8, Canada; pgmartin@cita.utoronto.ca}

\begin{abstract}
We present the results of \spi\ mid-infrared spectroscopic
observations of two highly-obscured massive X-ray binaries: \igrj\ and
\gx.  Our observations reveal for the first time the extremely rich
mid-infrared environments of this type of source, including multiple
continuum emission components (a hot component with $T$ $>$ 700~K and
a warm component with $T$ $\sim$ 180~K) with apparent silicate
absorption features, numerous \ion{H}{1} recombination lines, many
forbidden ionic lines of low ionization potentials, and pure
rotational \hh\ lines.  This indicates that both sources have hot and
warm circumstellar dust, ionized stellar winds, extended low-density
ionized regions, and photo-dissociated regions.  
It appears difficult to attribute the total optical extinction
of both sources to the hot and warm dust components, which suggests
that there could be an otherwise observable colder dust component responsible
for the most of the optical extinction and silicate absorption features.
The observed mid-infrared
spectra are similar to those from Luminous Blue Variables, indicating
that the highly-obscured massive X-ray binaries may
represent a previously unknown evolutionary phase of X-ray binaries with
early-type optical companions.  Our results highlight the importance and utility
of mid-infrared spectroscopy to investigate highly-obscured X-ray
binaries.
\end{abstract}

\keywords{circumstellar matter --- infrared: stars --- stars: emission-line, Be --- X-rays: binaries
--- X-rays: individual (\gx, \igrj)}

\section{Introduction} \label{sec_intro}

Recently, a large number of highly-obscured (e.g., \nh\ $\ge$
10$^{23}$ cm$^{-2}$) massive X-ray binaries have been discovered by
the \integ\ hard X-ray ($\ge$ 15 keV) satellite \citep{wet03}.  The
prototypical case is \igrj\ which shows variable high obscuration in
the X-ray, sometimes reaching \nh\ $\simeq$ 2 $\times$ 10$^{24}$
\citep{coet03, walteret03}.  Its bright optical and near-infrared (IR)
counterpart is an early B-type supergiant star with numerous emission
lines \citep{fc04}.  Interestingly the obscuration toward \igrj\
obtained in the optical and near-IR wavebands (\av\ $\sim$ 18) is
almost two orders of magnitude smaller than that inferred from the
X-rays, which suggests that
the extreme obscuration seen in the X-ray is intrinsic only to the
X-ray source.  However, \av\ $\sim$ 18 is still  greater than the
interstellar obscuration, indicating the existence of substantial
circumstellar material around the supergiant companion.

In order to fully understand the implications of the \integ\
discoveries --- specifically if they imply existence of a separate
class of highly-obscured X-ray binaries --- we must investigate any
similarities between the new \integ\ sources and previously known
sources.  As already suggested by \citet{ret03}, the X-ray pulsar \gx\
appears similar to the new \integ\ highly-obscured massive X-ray
binaries: it has variable X-ray obscuration of \nh\ $\simeq$
10$^{23}$--10$^{24}$ cm$^{-2}$, the compact source is a
neutron star, and the optical companion is an early B-type supergiant
\citep[or hypergiant;][]{ket95}, like \igrj.  Recently, \citet[][;
hereafter Paper~I]{kmr06}, using optical, near-, and mid-IR
($\le$~20~$\mu$m) spectral energy distributions (SEDs), have found that
both sources have strong mid-IR excesses that they identify as
continuum emission from hot dust.  This is suggestive that they have
very similar circumstellar material which may be related to their
strong X-ray obscuration.  In this {\em Letter}, we present the
results of \spi\ mid-IR spectroscopic observations of \igrj\ and \gx,
showing that both sources indeed have very similar rich mid-IR
properties previously unknown for X-ray binaries.

\section{Observations and Data Reduction} \label{sec_obs}

We observed \igrj\ and \gx\ with {\em InfraRed Spectrogrpah} (IRS)
aboard \spi\ \citep{jhet04, wet04} on 2005 September 8 and 2005 July
2, respectively.  The duration of the total observations was 1.32
(\igrj) and 0.82 (\gx) hours with the all six IRS modules in operation
in the standard staring mode, obtaining two nods of spectra separated
by 1/3 of the slit length.  For the basic data reduction, we used the
co-added Basic Calibration Data (BCD) produced through the standard
\spi\ Science Center data reduction pipeline (S13.2.0), and also the
\spi\ IRS Custom Extraction (i.e., SPICE) software (v1.3-beta1).  In
the low-resolution mode where the slit lengths are large (i.e.,
57\arcsec\ for 5.2--14.5 $\mu$m and 168\arcsec\ for 14.0--38.0
$\mu$m), we subtracted out background emission of one nod using the
spectrum of the another nod, and combined the two
background-subtracted spectra for the final spectrum.  However, in the
high-resolution mode, the small slit length (i.e., 11\farcs3 for
9.9--19.6 $\mu$m and 22\farcs3 for 18.7--37.2 $\mu$m), together with
the high brightnesses of the sources, made it impossible to subtract out
the background.  Therefore, we just combined the spectra from the two
nods for the final high-resolution spectrum without any background
subtraction. In addition, we used IRSCLEAN (v1.5) to mask and clean
rogue pixels from the co-added BCD images, and then interpolated over
neighboring normal pixels.

\section{Low-Resolution Spectrum: Continuum and Broad Features} \label{sec_lowres}

\begin{figure}[htf]
\plotone{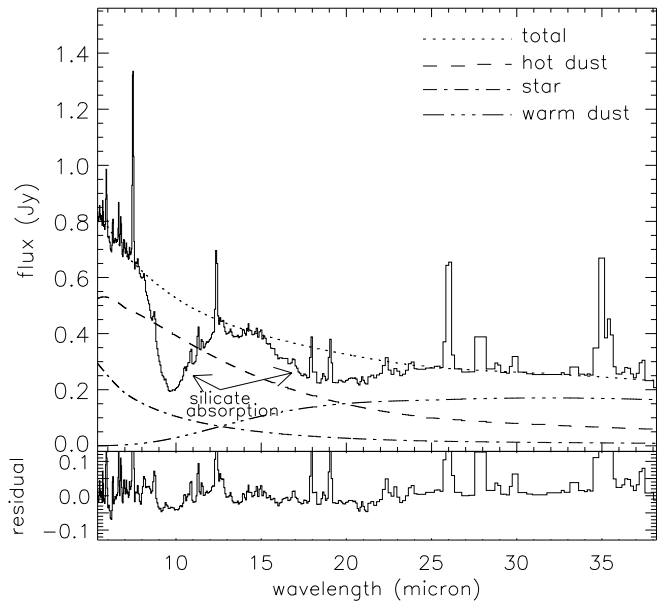}
\caption{{\em Main Panel:} The background-subtracted IRS low-resolution spectrum of \igrj.  
The solid line represents the observed spectrum, while the dotted line shows the
best-fit spectrum combining the hot dust (dashed line), the stellar
(dotted-dashed line), and the warm dust (dot-dashed line) components.
The SED fit was conducted excluding the wavelengths associated with the silicate absorption
features indicated by the arrows around 9.7 and 18 $\mu$m. 
{\em Bottom Panel:} The residuals of the SED fits. 
Note that for the residual the fluxes of the wavelengths associated with 
the silicate absorption features were calculated with the best-fit parameters
(see text). The unit of the residual is Jy.}
\label{fig:igrj}
\end{figure}

\begin{figure}[htf]
\plotone{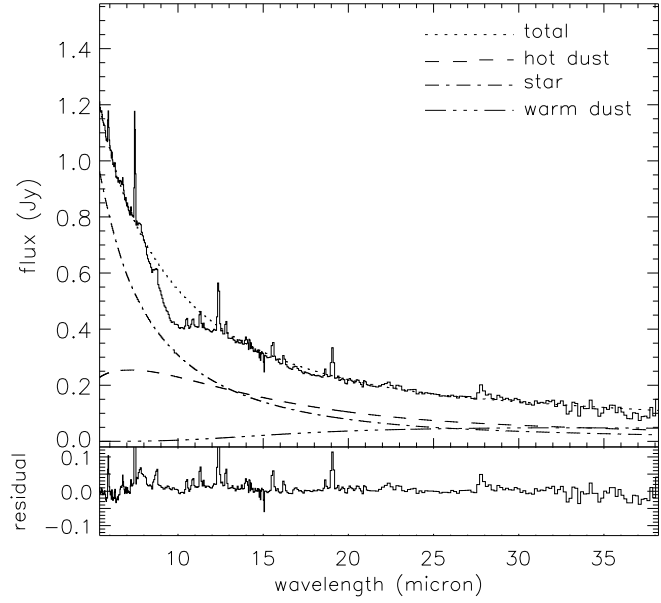}
\caption{Same as Figure~\ref{fig:igrj}, but for \gx.
The silicate absorption features are not as prominent as in \igrj,
but are still visible, especially around $\sim$9.7 $\mu$m.} 
\label{fig:gx}
\end{figure}

Some emission features,
such as ionic forbidden lines, \hh\ lines,
and Polycyclic Aromatic Hydrocarbon (PAH) emission,
present in the low-resolution spectra appear extended.
Here we consider just the emission from point sources, 
and defer discussion of the extended emission to a separate paper.
Figure~\ref{fig:igrj} presents the background-subtracted
low-resolution spectrum of \igrj, 
as well as the best-fit spectra of the continuum emission and the residual of the SED fit (see below).
Note that the low-resolution spectrum is extracted for the continuum source (i.e., \igrj)
based on the standard point source extraction procedure of SPICE.
Several features are prominent in Figure~\ref{fig:igrj}, 
including complicated multi-component continuum emission,
broad and strong absorption around 9.7 and 18 $\mu$m, and numerous
emission lines.  The spectrum of \gx\ (Figure~\ref{fig:gx}) is similar,
with strong  continuum emission,  silicate
absorptions feature around 9.7 $\mu$m, and several emission lines.

As in Figures~\ref{fig:igrj} and \ref{fig:gx},
the silicate absorption features and numerous emission lines 
make it difficult to carry out any meaningful independent SED fits
of the shorter wavelength (i.e., $\lambda$ $\le$ 8 $\mu$m) data of the IRS spectra.
Therefore, instead, we first examined if the IRS SEDs at the shorter wavelengths
are consistent with what have been found in Paper~I using only the line-free regions.  
According to Paper~I, the SEDs of \igrj\ and \gx\ at $\lambda$~$\le$~8~$\mu$m have 
contributions from hot (i.e.,T $\simeq$ 1040~K for \igrj\ and 720~K for \gx) dust emission and
stellar emission of a B0I star with temperature of 26000~K,
while the contribution of free-free emission is negligible.
Our IRS SEDs of the line-free regions in Figures~\ref{fig:igrj} and
\ref{fig:gx} at $\lambda$~$\le$~8$\mu$m were fitted very well with
those parameters, resulting in the reduced chi-square $\chi^2_\nu$
$\le$ 1.  We used the extinction \av\ = 18.5 and 7.1 for \igrj\ and
\gx, respectively, as in Paper~I.  However, in the longer wavelength
(i.e., $\lambda$~$\ge$~22~$\mu$m) range, we were unable to fit the IRS
SEDs with the given hot dust and stellar parameters because the best
fit resulted in unacceptably large chi-square values.  We therefore
performed SED fits with three components (i.e., hot dust, stellar, and
warm dust component) using the entire line-free SEDs in the
$\lambda$~$\le$~8~$\mu$m and $\ge$~22~$\mu$m range.  Over the fits the
parameters of the warm dust component were treated as free parameters,
while those of the hot dust and stellar component were fixed to be the
values in Paper~I.  As a result, we obtained the warm dust
temperature of 190~K (\igrj) and 170~K (\gx) with $\chi^2_\nu$
$\simeq$ 1.2 (\igrj) and 1.5 (\gx).  We used the extinction
coefficients compiled in \citet{m2000} excluding the 8--22 $\mu$m
range in order to avoid the uncertainties of the coefficients
associated with the silicate absorption features.  We instead
interpolated the extinction coefficients at $\lambda$ $\le$ 8 and
$\ge$ 22 $\mu$m to calculate the silicate absorption-free extinction
coefficients in the $\lambda$ = 8--22 $\mu$m range, and used them in the fits.
The main panel of Figures~\ref{fig:igrj} and \ref{fig:gx} presents the results of
our best SED fit obtained by excluding the silicate absorption features,
where we can confirm that the fit matches nicely 
the observed SED except the $\lambda$ = 8--22 $\mu$m range.
In the bottom panel, on the other hand,
we show the redisuals of the fit obtained by subtracting 
the fluxes computed by the best fit from the observed values.
For this residual construction, in order to see if our best SED fit 
is consistent with the observed silicate absorption features,
we combined the best-fit parameters obtained at $\lambda$ $\le$ 8 and $\ge$ 22 $\mu$m 
with the extinction coefficients in the $\lambda$ = 8--22 $\mu$m range \citep{m2000},
and calculated the silicate absorption feature-associated fluxes expected by the best SED fit
in the $\lambda$ = 8--22 $\mu$m range.
The residuals in Figures~\ref{fig:igrj} and \ref{fig:gx} show
that overall our best SED fit agrees with the silicate absorption features reasonably well,
although there appears to be some small discrepancies for \igrj.

The masses of the dust components were estimated under the
optically thin assumption \citep{het77}: $M_{\rm dust} =
F_{\lambda}d^2 / \kappa_{\lambda}B_{\lambda}(T_{\rm dust})$, where $d$
is the distance to the source, $\kappa_{\lambda}$ is the dust mass
absorption coefficient, and $B_\lambda(T)$ is the Planck function at
temperature $T$.  Using the interstellar mass absorption coefficients
of \citet{draine03} at 8 and 25 $\mu$m, the hot and warm dust masses
are estimated to be $M_{hot}$ $\simeq$ 4.5 $\times$ 10$^{-9}$ $d_5^2$ \msun\
and $M_{warm}$ $\simeq$ 3.1 $\times$ 10$^{-7}$ $d_5^2$ \msun\
for \igrj; $M_{hot}$ $\simeq$ 5.7 $\times$ 10$^{-9}$ $d_5^2$ \msun\ and
$M_{warm}$ $\simeq$ 1.4 $\times$ 10$^{-7}$ $d_5^2$ \msun\ for \gx, 
where $d_5$ is the distance to the sources normalized by 5 kpc. 

The silicate absorption features seen in Figures~\ref{fig:igrj} and
\ref{fig:gx} have a tight linear correlation with the optical
extinction in the diffuse interstellar medium (ISM),
although the correlation has been reported to break down 
in dense enviornment where the silicate feature does not 
show a monotonic increase with extinction 
at \av\ $\geq$ 12 mag \citep[e.g.,][and references therein]{cet07}.
The correlation in the diffuse ISM is \av/$\tau_{\rm 9.7}$ = 18.5 
\citep[][and references therein; $\tau_{\rm 9.7}$ is the optical depth of the silicate absorption at 9.7 $\mu$m]
{draine03}.  
If we adopt the above correlation given in the diffuse ISM,
the intensity ratio of the observed spectrum to the unabsorbed spectrum 
in Figures~\ref{fig:igrj} and \ref{fig:gx} at 9.7 $\mu$m 
correspond to \av\ $\sim$~18.5 and $\sim$~6 for \igrj\ and \gx, respectively, 
These are very similar to the values obtained in previous studies
for the total optical extinctions of the both sources.
This suggests that a substantial portion, if not all, 
of the silicate features seen in Figures~\ref{fig:igrj} 
and \ref{fig:gx} are associated with the optical extinction of \igrj\ and \gx.

\section{High-Resolution Spectrum: Line Emission} \label{sec_highres}

\begin{figure}[htf]
\plotone{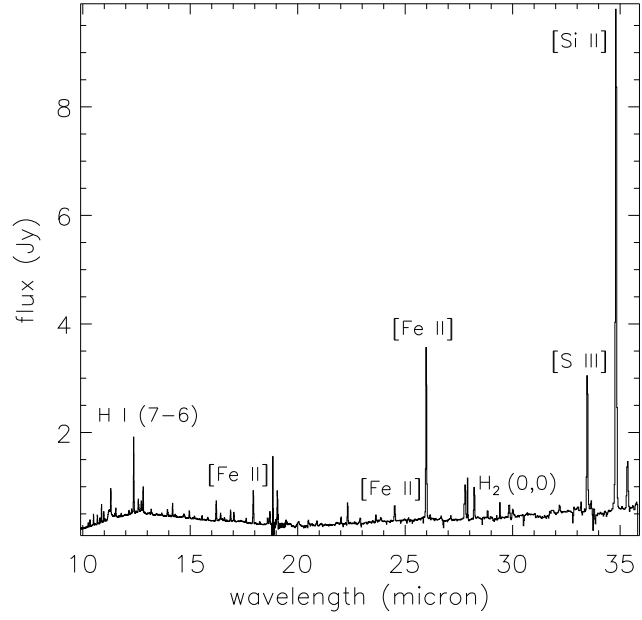}
\caption{The IRS high-resolution spectrum of \igrj.
Note that the background is {\em NOT} subtracted.
Some strong lines are marked.}
\label{fig:igrjhigh}
\end{figure}

\begin{figure}[htf]
\plotone{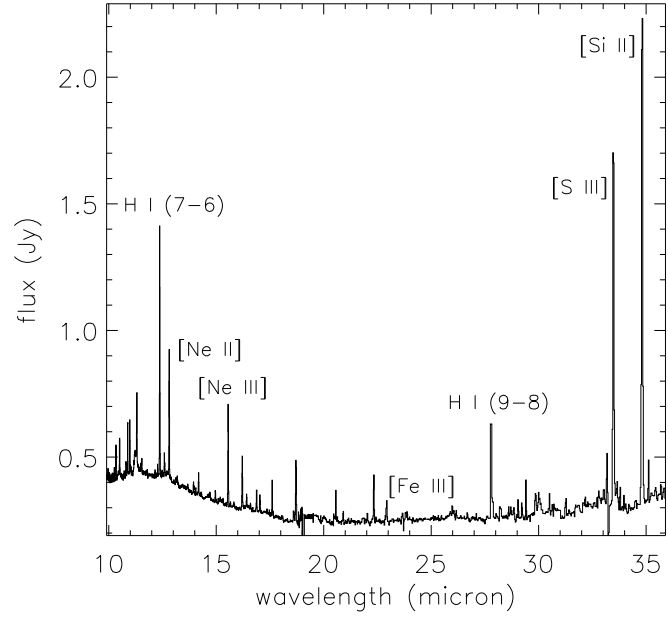}
\caption{Same as Figure~\ref{fig:igrjhigh}, but for \gx.}
\label{fig:gxhigh}
\end{figure}

Figures~\ref{fig:igrjhigh} and \ref{fig:gxhigh} present high-resolution
spectrum of \igrj\ and \gx, revealing numerous emission lines.  
(Note that the background emission is {\it NOT} subtracted.)
The majority of them are \ion{H}{1} lines, while the
rest consists of metallic forbidden lines, pure rotational \hh\ lines,
and 11.3 $\mu$m PAH emission feature.
Table~\ref{tbl_HI} lists the intensities of some of the bright
\ion{H}{1} lines identified in both sources with high ($>$ 3)
signal-to-noise ratio.  The line intensities and the signal-to-noise
ratio were estimated with Gaussian profile fits.  The intensities of
the \ion{H}{1} lines are consistent with gas at $T$ $\simeq$ 10$^4$ K
and with \necubic\ $\simeq$ 10$^4$ \cmcubic\ in the Menzel Case~B
state, following \citet{fc04}, although the intensity ratios are not
very sensitive to those parameters \citep{hs87}.

Table~\ref{tbl_other} lists the parameters of forbidden ionic lines and \hh\ lines.
For the forbidden lines, \neii, \neiii, \siii, and \silii\ were detected
in both sources, while \niii\ and \feii\ were only in \igrj; \feiii\
is only in \gx. 
The \hh\ lines were detected in both sources.  The intensity ratio of
the \siii\ lines of 18.7 and 33.5 $\mu$m is diagnostic of the number
density in the temperature range of 5000--20000~K.  The
extinction-corrected \siii\ line surface brightness ratio of 18.7/33.5
$\mu$m is 1.21 for \igrj\ and 1.68 for \gx: according to the calculation
of \citet{het84} and \citet{aet99}, the ratios correspond to the
electron number density \necubic\ $\sim$ 1000 \cmcubic\ for both
sources.  On the other hand, the intensity ratio of the \neii\ line
(12.8 $\mu$m) and the \neiii\ line (15.6 $\mu$m) is sensitive to the
hardness of the radiation, or temperature of the environment.  If we
use the calculation of \citet{ket96}, the extinction-corrected
intensity ratio of \neii/\neiii\ is consistent with $\sim$
4~$\times$~10$^4$~K for \igrj\ and $\sim$ 4.5~$\times$~10$^4$~K for
\gx.  The three \hh\ lines detected in the both sources are pure
rotational transition lines of \hh~(0,0)~S(0,1,2) which are usually
excited by collisions.  Assuming that the bottom two lines follow the
distribution of the local thermodynamic equilibrium and are optically
thin, the surface brightness ratios are consistent with the excitation
temperature $T_{\rm ex}$ $\sim$ 1000~K.
The estimated \hh\ column densities are $\sim$ 1 $\times$ 10$^{19}$
cm$^{-2}$ (\igrj) and 5 $\times$ 10$^{18}$ cm$^{-2}$ (\gx).  This
implies that the optical extinction associated with the warm molecular
gas is very small, \av\ $\ll$ 1, based on the relation \nh\ = 1.87
$\times$ 10$^{21}$ \av\ \citep{draine03}.

\section{Discussion and Conclusion} \label{sec_dis_sum}

Our \spi\ spectroscopic observations of \igrj\ and \gx\ have revealed
for the first time the rich mid-IR environment of highly-obscured X-ray binaries.
This includes two dust components with prominent silicate absorption,
numerous \ion{H}{1} recombination lines, many forbidden ionic  lines,
and pure rotational \hh\ lines.
Based on the observed spectra, we infer the following components for
\igrj\ and \gx: (1) hot ($T$ $>$ 700~K) and warm ($T$ $\sim$ 180~K)
circumstellar dust; (2) ionized stellar winds responsible for the
\ion{H}{1} lines; (3) extended low-density ionized regions for the
forbidden lines; and (4) photo-dissociated regions associated with the
PAH, \hh\ and possibly the \silii\ line emission.  For the forbidden
lines, all the detected lines have relatively low ionization potentials
like in starburst galaxies where the radiation is relatively soft (compared with active
galactic nuclei, for instance).  This may indicate that the
illumination of hard X-rays from the central compact X-ray source is
not primarily responsible for the forbidden line emission.  However,
the inferred temperature for the radiation exciting the \neii\ and
\neiii\ lines is hotter than the stellar photospheres, so there can
be some contribution from the compact object.
Considering that \niii\ and \feii\ were detected only in \igrj\ while
\feiii\ was only in \gx, the radiation field of \igrj\ may be
softer than \gx, as demonstrated by the small temperature difference between
the two sources (see \S~\ref{sec_highres}).

Perhaps the most natural explanation for the origin of the hot dust
component relies on dust formation in the dense outflows from the
early-type companions of \igrj\ and \gx, as is the case in most sgB[e]
stars (i.e., B-type supergiants with forbidden emission lines).
However, the origin of the warm circumstellar dust component is very uncertain, 
and B[e] stars seldom show evidence for this type of warm dust.  
If both the hot and warm dust components have spherical shell geometres around the central star,
then the associated optical extinctions are:
\av\ $\simeq$ 4 $\times$ 10$^{-4}$ $Q_{\rm abs}$ ($M_{\rm d}$/10$^{-6}$ \msun)
($T_{\rm d}$/100 K)$^4$ ($L_{\rm UV}$/10$^{39}$ ergs s$^{-1}$),
where $M_{\rm d}$ and $T_{\rm d}$ are the mass and temperature of the dust components,
$Q_{\rm abs}$ $<$ 1 is the dust absorption coefficient, 
and $L_{\rm UV}$ is the ultra-violet luminosity of the central star.
This gives the optical extinctions \av\ $\ll$ 1 for both the hot and warm dust components
of \igrj\ and \gx. (Here we use 1 $\times$ 10$^{39}$ ergs s$^{-1}$ for $L_{\rm UV}$ for both sources.)
Therefore, under the assumption of spherical shell geometry, 
both the hot and warm dust components of \igrj\ and \gx\ contribute
very little to the total optical extinction, suggesting that
the hot and warm dust components are {\em NOT} strongly associated with 
the silicate absorption features.
This also applies to the dust associated with the warm extended \hh\ gas 
since the optical extinction from this component is tiny 
(i.e., Av $\ll$ 1; see \S~\ref{sec_highres}). 
What's then the origin of the silicate absorption features?
If it is due to the foreground ISM, we would expect to see the 
$\rm CO_2$ ice feature around 15 $\mu$m, especially for \igrj, based on the intensity ratio
between the silicate absorption feature and ice feature found in the ISM \citep{ket05}. 
The absence of the ice feature in our spectra
supports the interpretation that the silicate absorption features are probably
not associated with the foreground ISM.
One possibility may be the existence of an undisclosed colder (e.g., $\ll$ 100~K) 
circumstellar dust component which is responsible for the silicate absorption features
and most of the optical extinction.
We need further longer wavelength observatons to confirm this possibility.
Considering that the 9.7~$\mu$m
silicate absorption feature represents the oxygen-rich material, the
existence of the 9.7 micron silicate absorption feature 
may indicate that the origin of the potential colder dust component 
is related to the nucleosynthesis of the progenitors of \igrj\ and \gx\
(or their companions).

The optical/near-IR companion of \igrj\ is a sgB[e] star.  Such stars
are known to have hot ($T$ $\sim$ 1000~K) circumstellar dust, and
probably are evolving into Luminous Blue Variables (LBVs) or
Wolf-Rayet stars.  Our mid-IR spectra of \igrj\ and \gx\ are very
similar to that of the LBV P~Cygni which shows many \ion{H}{1} lines
and forbidden ionic lines \citep{let96b}.  The difference is that
while the mid-IR continuum of P~Cygni is due to the free-free emission
in stellar winds, the main mid-IR emission of \igrj\ and \gx\ is from
multiple dust continua.  (The SEDs of both sources observed here are
inconsistent with the free-free emission, as mentioned in Paper~I.)
However, we note that some LBV stars have also been observed to have
thermal mid-IR dust emission \citep{let96a}. Based on the fact that B[e]
stars seldom show mid-IR forbidden line emission, the B-type
supergiant (or hypergiant) companions of \igrj\ and \gx\ may be in the
evolutionary track to LBVs, implying that the extremely high
obscuration seen in some massive X-ray binaries may be a phenomenon
associated with the evolutionary phase.  This scenario is also consistent
with the fact that the highly-obscured massive X-ray binary Cygnus~X-3
has a Wolf-Rayet star companion, together with the circumstellar dust
emission of $T$ $\sim$ 250~K \citep{ket02}.  

\acknowledgments
This work is based on observations made with the Spitzer Space Telescope, 
which is operated by the Jet Propulsion Laboratory, 
California Institute of Technology, under a contract with NASA.
D.-S.M thanks Elise Furlan for her help in IRS data analysis 
and Marten van Kerkwijk for comments.
This research was partly supported by the Discovery
Grant (327277) of Natural Science and Engineering Research Council of Canada to D.-S.M.

\clearpage
\begin{deluxetable}{crr||crr}
\tablecolumns{6}
\tablewidth{0pt}
\tablecaption{Observed Intensities  of \ion{H}{1} Transitions\label{tbl_HI}}
\tablehead{
\colhead{Transition($\mu$m)} & \colhead{IGR}    & \colhead{GX}     & \colhead{Transition($\mu$m)} & \colhead{IGR}    & \colhead{GX}  \\
\colhead{}     & \colhead{J16318} & \colhead{301--2} & \colhead{}     & \colhead{J16318} & \colhead{301--2}}
\startdata
17--9(10.26)   &   4.39           &  2.44            &  12--8(10.50)   & 11.57            &  11.40 \\ 
16--9(10.80)   &   5.60           &  2.20            &   9--7(11.31)   & 25.43            &  13.50 \\ 
22--10(11.49)  &   1.74           &  0.78            &  15--9(11.54)   &  6.48            &   1.87 \\ 
20--10(12.16)  &   2.84           &  1.35            &   7--6(12.37)\tablenotemark{a}   & 72.02            &  42.12 \\ 
18--10(13.19)  &   5.15           &  1.70            &  13--9(14.18)   & 10.61            &   3.37 \\ 
23--11(14.30)  &   2.00           &  0.93            & 22--11(14.71)   &  5.02            &   0.87 \\ 
16--10(14.96)  &   5.41           &  1.75            & 20--11(15.82)   &  2.79            &   1.33 \\ 
10--8(16.21)   &  14.66           &  7.32            & 15--10(16.41)   &  3.83            &   1.83 \\ 
19--11(16.59)  &   2.15           &  0.40            &  12--9(16.88)   &  7.35            &   3.17 \\ 
14--10(18.62)  &   4.04           &  1.38            & 16--11(20.92)   &  2.69            &   1.14 \\ 
9--8(27.80)    &  18.71           &  8.86            &  \nodata        &  \nodata         &  \nodata   \\ 
\enddata
\tablecomments{The values are in 10$^{-14}$ ergs s$^{-1}$ cm$^{-2}$ and are not corrected for extinction.}
\tablenotetext{a}{The \ion{H}{1} 11--8 transitions may be blended.}
\end{deluxetable}

\clearpage
\begin{deluxetable}{crr||crr}
\tablecolumns{6}
\tablewidth{0pt}
\tablecaption{Observed Intensities of Other Lines \label{tbl_other}}
\tablehead{
\colhead{Line($\mu$m)} & \colhead{IGR}    & \colhead{GX}     & \colhead{Line($\mu$m)} & \colhead{IGR}    & \colhead{GX}  \\
\colhead{}     & \colhead{J16318} & \colhead{301--2} & \colhead{}     & \colhead{J16318} & \colhead{301--2}}
\startdata
\niii(10.68)   &  10.25        &  \nodata  & \niii(12.73)   &  13.33        &  \nodata      \\ 
\niii(18.24)   &   1.69        &  \nodata  & \nodata        &  \nodata      &  \nodata      \\ 
\neii(12.81)   &  21.44        &  23.41    & \neiii(15.56)  &  2.60         &     15.67     \\ 
\feii(17.94)   &  22.54        & \nodata   & \feii(24.52)   &  7.33         &  \nodata      \\
\feii(25.99)   &  70.93        & \nodata   & \feii(35.35)   &  15.46        &  \nodata      \\
\feiii(22.93)  &  \nodata      &   1.99    & \nodata        & \nodata       &  \nodata      \\
\siii(18.71)   &   7.73        &   8.02    & \siii(33.48)   &  40.14        &     25.01     \\
\silii(34.82)  &  177.20       &  33.33    & \nodata        & \nodata       &  \nodata      \\
\hh(0,0)S(2)(12.28) &  2.88    &  1.44     & \hh(0,0)S(1)(17.03) &  4.66         &    2.05       \\
\hh(0,0)S(0)(28.22) & 10.08    &  4.24     & \nodata             &  \nodata      &  \nodata      \\
\enddata
\tablecomments{The values are in 10$^{-14}$ ergs s$^{-1}$ cm$^{-2}$ and are not corrected for extinction.}
\end{deluxetable}

\clearpage

\clearpage

\clearpage

\clearpage

\end{document}